\documentclass[twocolumn,twoside,slac_two]{revtex4}
\usepackage{graphicx}
\usepackage{fancyhdr}
\usepackage{hyperref}
\hypersetup{
    colorlinks=true,
    urlcolor=magenta,
}
\begin{document}

\title{Forbush Decreases during the DeepMin and MiniMax of Solar Cycle 24}


\author{D. Lingri, H. Mavromichalaki}
\affiliation{Athens Cosmic Ray Group, Faculty of Physics, National and Kapodistrian University of Athens, 15784, Athens, Greece}
\author{A. Belov, E. Eroshenko, V. Yanke, A. Abunin, M. Abunina}
\affiliation{Pushkov Institute of IZMIRAN RAS, Moscow, Russia}

\begin{abstract}
After a prolong and deep solar minimum at the end of solar cycle 23, the current cycle 24 is one of the lowest cycles.
The two periods of deep minimum and mini-maximum of the cycle 24 are connected by a period of increasing solar activity. In this work, the Forbush decreases of cosmic ray intensity during the period from January 2008 to December 2014 are studied. A statistical
analysis of 749 events using the IZMIRAN database of Forbush effects obtained by processing the data of the worldwide
neutron monitor network using the global survey method is performed. A further study of the events that happened on the Sun and affected the interplanetary space, and finally provoked the decreases of the galactic cosmic rays near Earth
is performed. A statistical analysis of the amplitude of the cosmic ray decreases with solar and geomagnetic parameters is
carried out. The results will be useful for space weather studies and especially for Forbush decreases forecasting.
\end{abstract}

\maketitle

\thispagestyle{fancy}


\section{INTRODUCTION}
Cosmic rays (CR) are high energetic particles that are coming from galactic and extragalactic sources \cite{Cane 2000}. They are not stable but a lot of cosmic ray variations are recorded by the neutron monitor worldwide network that is considered as a reliable network of ground based detectors of cosmic rays. These variations are a really important source on space weather information.

The first who studied the cosmic ray variations almost eighty years, or otherwise eight solar cycles ago, was Scott Forbush \cite{Forbush 1954}. He discovered a very important variation in cosmic ray intensity, after whom it was named as Forbush decrease (FD). In this work we study FDs with amplitude greater than 2\%. 
As a source of FD is considered an ICME, which have created an interplanetary shock, and when it reaches the Earth$' $s magnetosphere, in the most cases, a sudden storm commencement (SSC) is created and a decrease in cosmic rays appears \cite{Belov 2008}. 

Previous studies \cite{Papailiou 2012a} \cite{Papailiou 2012b} have introduced the meaning of FDs’ precursors and the effort of making from them a forecasting model for FDs. As a precursor is considered a pre - decrease within the narrow longitude range and a wide pre - increase. 

In this work we have studied the FDs of the cosmic ray intensity during the time period from January 2008 to December 2014 and we have compared them with the FDs which had been taken place in the previous solar cycle 23. Furthermore we present preliminary results of FDs precursors in some events that have been selected with specific criteria. 

\section{Data Selection}
About 50 Neutron Monitors located all over the world provide their cosmic ray data to the High Resolution Neutron Monitor Database - NMDB (http://www.nmdb.eu). In this work corrected for pressure hourly values of these data have been obtained. In addition the IZMIRAN database of FDs has also used extracting a lot of information for each selected event. By applying the Global Survey Method (GSM), the first harmonic of the cosmic ray anisotropy at a cosmic ray rigidity of 10GV, which is close to the effective rigidity of the particles being registered by the neutron monitor worldwide network, is calculated \cite{Belov 2005}.

\begin{figure}[t]
\centering
\includegraphics[width=90mm]{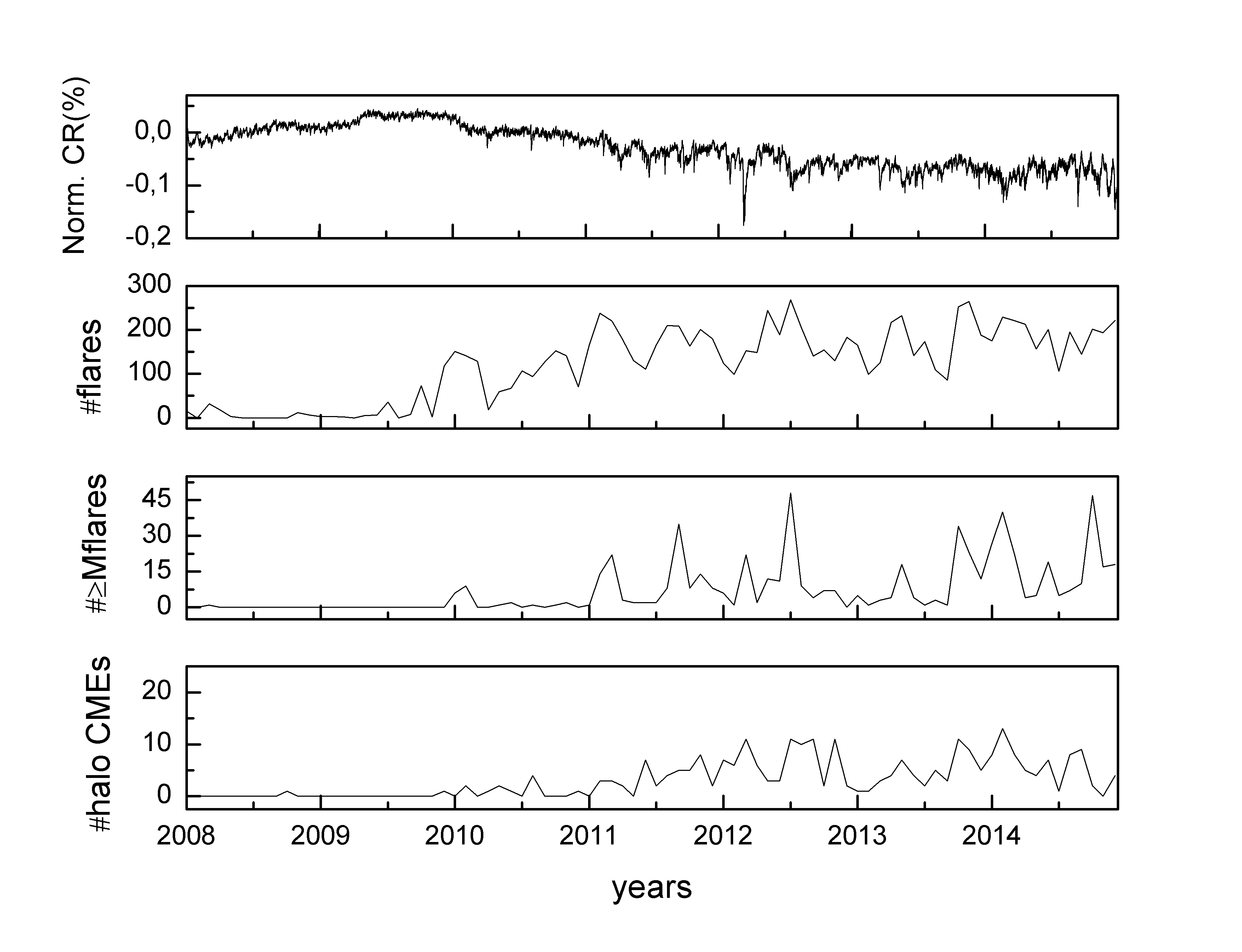}
\caption{Time profiles of the cosmic ray intensity (upper panel), the solar flares (second panel), the solar flares of $\geq$ M class (third panel) and halo CMEs (bottom panel)\label{fig1}} 
\end{figure}

The time period from January 2008 to December 2014, covering the minimum bewteen the solar cycles 23/24, the ascending phase and the maximum period of the solar cycle 24, was studied. In Figure ~\ref{fig1}, from top to bottom, time profiles of the monthly values of the normalized cosmic ray intensity from the Fort Smith Neutron Monitor, the total number of solar flares and the number of importance $\geq$ M flares from the GOES satellites (ftp.ngdc.noaa.gov) are presented. In the bottom panel of this Figure the number of halo CMEs from the Large Angle and Spectroscopic Coronagraph (LASCO) onboard the Solar and Heliospheric Observatory (SOHO) (http://cdaw.gsfc.nasa.gov) are also illustrated.

An extended and deep minimum between the solar cycles 23 and 24, where it can be said that there was no solar activity, can be observed. There were only three halo CMEs in about two years. We have to notice here that the cosmic ray intensity in this time period was in the highest levels that have ever been recorded. In the next two years, 2010-2011, the solar activity started to increase and the cosmic ray intensity, on the contrary, started to decrease. So these two years are characterized as the ascending phase of solar cycle 24. Then in 2012 the solar activity developed a shoulder, which turned into a plateau in 2013 and had an extra peak in 2014, in a good agreement with the results of other studies (e.g. \cite{Aslam 2015}).

\begin{figure}[t]
\centering
\includegraphics[width=75mm]{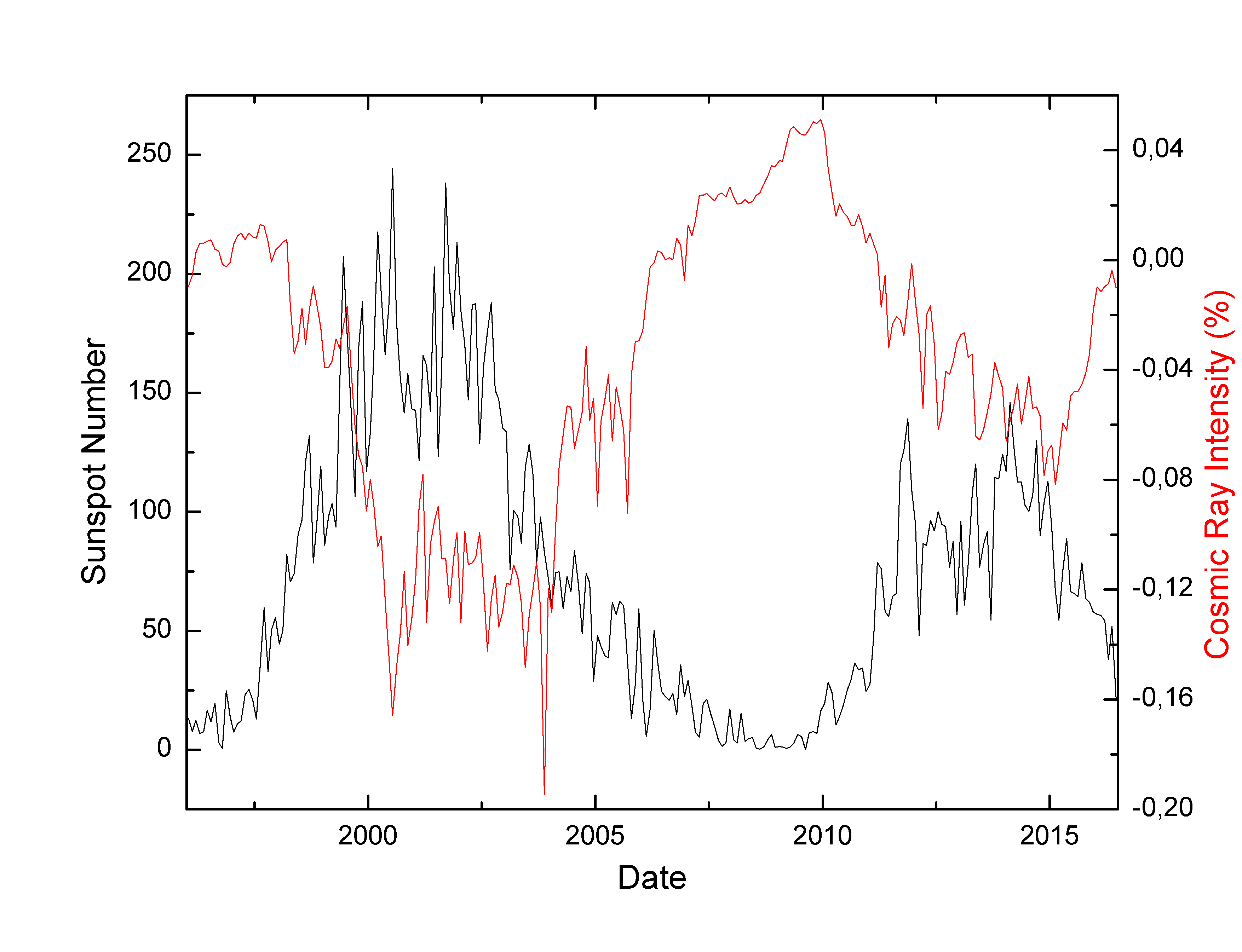}
\caption{The anti-correlation between the number of sunspots and the cosmic ray intensity for the last two solar cycles.\label{fig2}} 
\end{figure}

\begin{figure}[t]
\centering
\includegraphics[width=75mm]{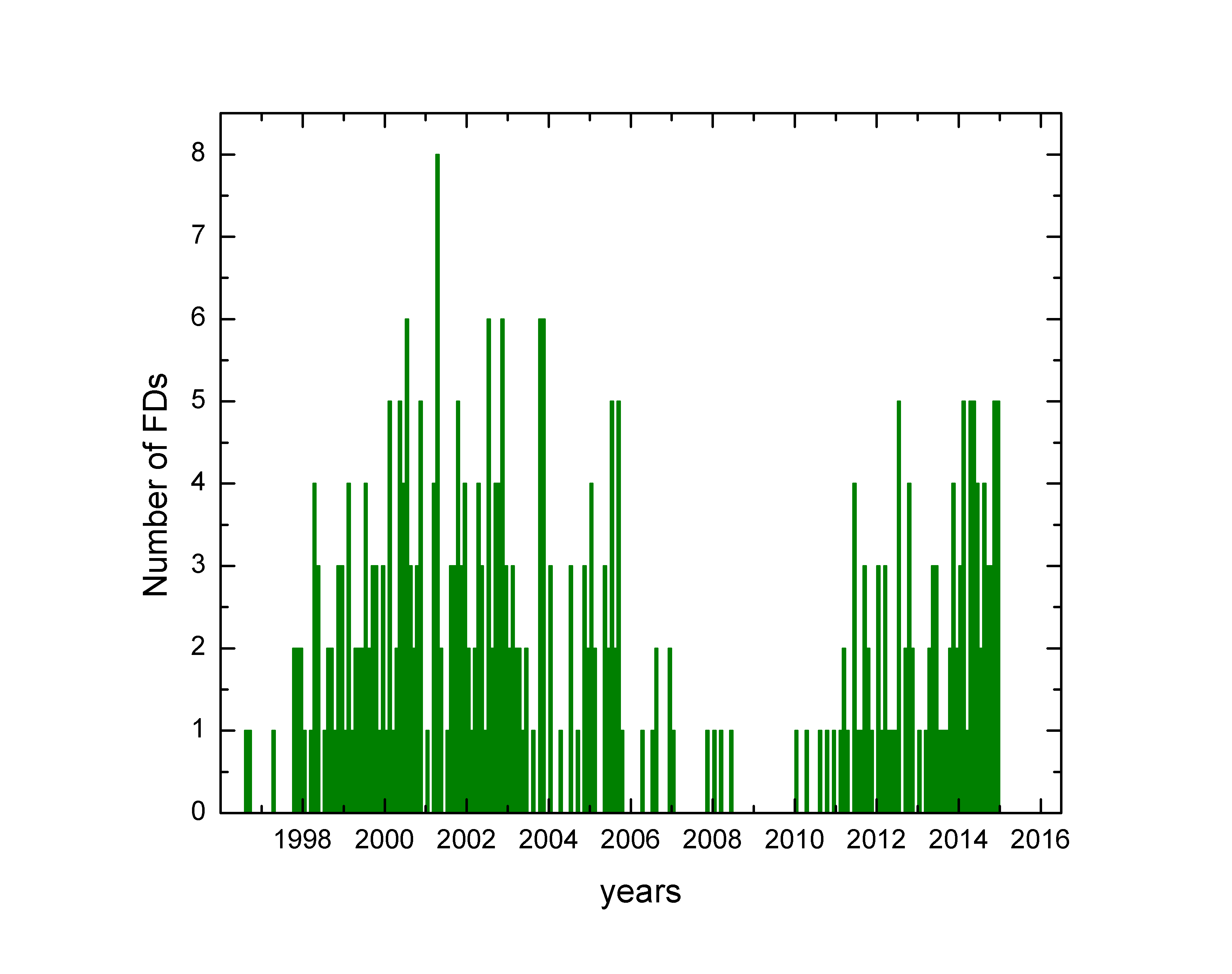}
\caption{The number of FDs per month during the two last solar cycles.\label{fig3}} 
\end{figure}

The anti-correlation between the number of sunspots and the cosmic ray intensity during the last two solar cycles is well noticed in Figure \ref{fig2}. The deep minimum area and the minimax area of solar cycle 24 are obvious. If we have a look on the number of FDs per month (Figure ~\ref{fig3}) that occurred during these solar cycles, it can be noted that more FDs took place at the descending phase of solar cycle 23 and the same seems to be happened also at the current cycle. On the ascending phases only a few FDs were recorded. A few events were also observed at the peak of the current cycle, with a maximum number of five events per month.

\section{Data analysis}
During the whole studied period of cycle 24 a number of 749 Forbush Effects recorded, out of which only 114 events with amplitude greater than 2\%. From these events, only these that had amplitude greater than 5\% are presented in Table \ref{l2ea4-t1}. Some of their characteristics are the amplitude of the decrease of the cosmic ray intensity, the values of the geomagnetic index Dst and of the interplanetary magnetic field (IMF), the solar wind velocity and the connected flares and CMEs, as also the CMEs velocity (see also \cite{Lingri 2016}). It is noted that from the cases of fifteen FDs with amplitude greater than 5\% eight of them occurred in 2014 while the other seven events occurred in a time period of six years. 

It can be said that there is a quasi - linear correlation between geomagnetic index and the FD’s amplitude \cite{Lingri 2016}. But it is not a rule that a large value of a geomagnetic index, in the case of Dst, automatically means a small FD, as it is presented in Table \ref{l2ea4-t1}. All FDs in Table \ref{l2ea4-t1} are created by CMEs. It is naturally that a CME is the only source of a big FD \cite{Belov 2008}.

The majority of FDs are produced by CMEs with velocities from 400 to 1200 km/sec and at the same time the velocity of the solar wind fluctuates from 400 to 740 km/sec. On the contrary, once again it is resulted that the FD$' $s amplitude is not tightly connected with the CMEs velocity and the same is valid for the connection of them with the solar wind velocity \cite{Lingri 2016}.

\begin{table*}[t]
\begin{center}
\caption{List of FDs with amplitude greater than 5\%}
\begin{tabular}{|l|c|c|c|c|c|c|c|c|c|}
\hline \textbf{} & \textbf{SSC} & \textbf{Ampl.} &
\textbf{Dst min}&\textbf{IMF}&\textbf{Vsw}&\textbf{}&\textbf{Date of CME occur.}&\textbf{Vcmes}\\
 \textbf{A/A} & \textbf{DD.MM.YYYY} & \textbf{CR} &
\textbf{(nT)}&\textbf{(nT)}&\textbf{(km/s)}&\textbf{Flares}&\textbf{DD.MM.YYYY}&\textbf{(km/s)}\\
 \textbf{} & \textbf{ hh:mm  UT} & \textbf{10GV (\%)} &
\textbf{}&\textbf{}&\textbf{}&\textbf{}&\textbf{ hh:mm:ss UT}&\textbf{}
\\
\hline  1 & 18.02.2011, 01:36 & 5.2 & -30 & 30.6 & 691 & X2.2 & 15.02.2011, 02:24:05 & 669\\
\hline  2 & 08.03.2012, 11:05 & 11.7 & -143 & 23.1 & 737 & X5.4 & 07.03.2012, 00:24:06 & 2684\\
\hline  3 & 12.03.2012, 09:21 & 5.7 & -51 & 23.6 & 727 & M8.4 & 10.03.2012, 18:12:06 & 1296\\
\hline  4 & 14.07.2012, 18:11 & 6.4 & -133 & 27.3 & 667 & X1.4 & 12.07.2012, 16:48:05 & 885\\
\hline  5 & 13.04.2013, 05:59 & 5.3 & -7 & 12.9 & 516 & M6.5 & 11.04.2013, 07:24:06 & 861\\
\hline  6 & 23.06.2013, 04:26 & 5.9 & -49 & 7.6 & 697 & M2.9 & 21.06.2013, 03:12:09 & 1900\\
\hline  7 & 14.12.2013, 14:00 & 5.1 & -41 & 10.9 & 600 & C4.6 & 12.12.2013, 03:36:05 & 1002\\
\hline  8 & 27.02.2014, 16:50 & 5.1 & -99 & 16.6 & 483 & X4.9 & 25.02.2014, 01:25:50 & 2147\\
\hline  9 & 18.04.2014, 02:00 & 5.1 & -13 & 10.2 & 506 & M1.0 & 16.04.2014, 20:00:05 & 764\\
\hline  10 & 16.06.2014, 08:00 & 5.4 & -14 & 7.2 & 393 & M1.1 & 15.06.2014, 13:00:05 & 958\\
\hline  11 & 11.09.2014, 23:45 & 8.1 & -16 & 14.0 & 467 & M4.5 & 09.09.2014, 00:06:26 & 920\\
\hline  12 & 12.09.2014, 15:53 & 8.5 & -75 & 31.7 & 730 & X1.6 & 10.09.2014, 18:00:05 & 1267\\
\hline  13 & 10.11.2014, 02:20 & 6.9 & -57 & 19.4 & 509 & X1.6 & 07.11.2014, 18:08::34 & 795\\
\hline  14 & 01.12.2014, 05:00 & 5.9 & -25 & 14.0 & 592 & C2.1 & 30.11.2014, 12:24:05 & 939\\
\hline  15 & 21.12.2014, 19:11 & 10.8 & -51 & 16.5 & 429 & M8.7 & 17.12.2014, 05:00:05 & 587\\
\hline
\end{tabular}
\label{l2ea4-t1}
\end{center}
\end{table*}

\section{Forbush decreases precursors}
For studying the precursor effects in different Forbush events, the “Ring of Stations” (RS) method has been applied \cite{Asipenka 2009} to the hourly data of CR intensity recorded by the neutron monitor stations of the world wide network with cut off rigidity Rc $<$4GV and latitudes $<$70° \cite{Papailiou 2012a}.

We follow the criteria of the model as they have been proposed by \cite{Papailiou 2012a}, such as the anisotropy criterion. The events of FDs through the time period 2008-2014 have been studied in detail and from them have been selected the cases which had an amplitude greater than 3\% and the anisotropy before the shock$' $s arrival was Axy \textgreater 0.9\%. The chosen anisotropy can be considered as anomalous, since it exceeds the mean statistical value significantly \cite{Belov 2008}. Furthermore these events have to be associated with a SSC and the interplanetary environment has to be quiet or with small disturbances before the FD’s appearance. From 58 events that followed the first criterion only seven of them reached the anisotropy value of 0.9\% before the shock (it is usually \textless0.6\%) and present precursors signs. All of them have been connected with a CME. In this study the characteristic event of 17 March 2013 was chosen to be analyzed and to be presented.

\begin{figure}[t]
\centering
\includegraphics[width=80mm,  height=40mm]{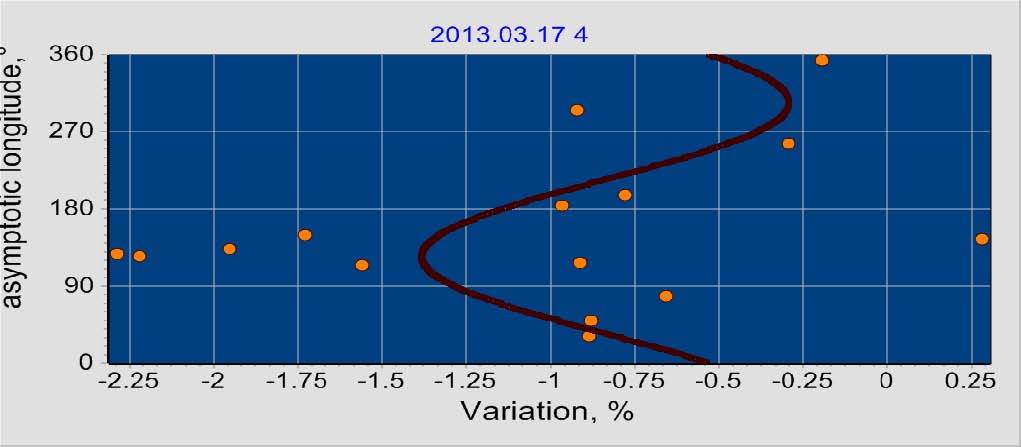}
\caption{Dependence of cosmic rays variations on the asymptotic longitude of neutron monitors in the event on 17 March 2013 at 4:00 (UT).\label{fig4-f4}} 
\end{figure}

\begin{figure}[t]
\centering
\includegraphics[width=80mm, height=40mm]{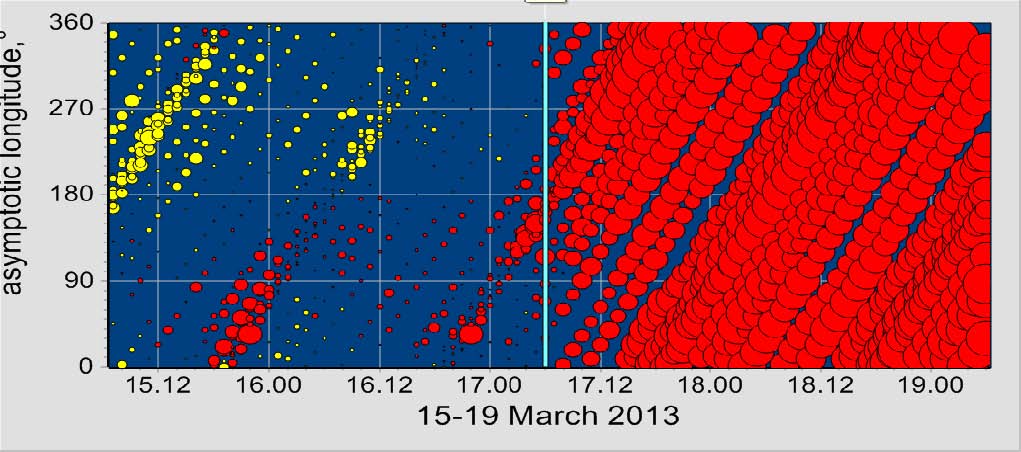}
\caption{The asympotic longitude - time distribution of the event of 17 March 2013.\label{fig5}} 
\end{figure}

This event of March 2013 was related to a CME registered on 15 March 2013 at 07:12:05UT, connected with a M1.1 flare (N09W05). The associated shock arrival was registered on 17 March 2013 at 05:59UT. There were strong changes in the interplanetary space parameters, especially in the interplanetary magnetic field intensity.

The amplitude of the occurred CR intensity decrease has been calculated to the value of 4.6\% at cosmic rays of 10GV by using the GSM method. In Figure \ref{fig4-f4}, orange dots are hourly cosmic ray variations at different neutron monitor stations, and the curve is the fitted first harmonic of anisotropy (solar diurnal variation) for data under consideration. There is a good example of pre-decrease as at a narrow region we can get deep decrease \textgreater2.25\%.

The asymptotic longitudinal CR distribution diagram for this event is presented in Figure \ref{fig5}. Here the CR intensity decreases, as measured by all neutron monitor stations used by the RS method, are depicted with red circles, while yellow circles refer to CR intensity increases relative to a quiet fixed period. The size of the circles is proportional to the size of the variation \cite{Papailiou 2012b}. The precursor is a pre-decrease in the longitudinal zone $0^{\circ} - 180^{\circ}$ before the shock$' $s arrival.

\section{conclusions}
Finally, to sum up, from 114 studied FDs with amplitude greater than 2\% at 10GV, only three occurred in the first two years (2008 - 2009), twenty one events were recorded during the ascending phase (2010 - 2011) and ninenty events in the minimax phase (2012 - 2014), with fourty five of them to be occurred in the last year 2014. From our analysis the following conclusions are obtaining:
\begin {itemize}

\item There is no obvious correlation of the FDs amplitude with the velocities of solar wind and of CMEs. 
\item There are signs of precursors that can help to forecast the upcoming Forbush decrease.
\item In the current solar cycle 24 only seven out of fifty eight FDs follow the anisotropy criterion, and only about 70\% of them present warning signals.
\item The pre-decreases and the pre-increases of the FDs occur on specific asymptotic longitudes. The pre-decrease of a FD of cosmic ray intensity is at a narrow region  and gives deep decreases with amplitude \textgreater2.25\%.

\end {itemize}

\bigskip % extra skip inserted
\begin{acknowledgments}
We acknowledge the NMDB database (www.nmdb.eu), founded under the European
Union’s FP7 Program (contract no. 213007) for providing high-resolution cosmic ray data and the PI of the neutron monitor stations. Thanks are due to ACE/Wind, OMNI, and NOAA data centers for kindly providing the related solar and interplanetary data.
Many thanks are also due to the cosmic ray group of the IZMIRAN of the Russian Academy of Sciences
for kindly providing the most complete catalog of Forbush decreases. D. Lingri thanks the National and Kapodistrian University of Athens and the Local Organizing Commitee of the XXV ECRS 2016 for supporting her participation to the XXV European Cosmic Ray Symposium. 
\end{acknowledgments}

\bigskip % extra skip inserted

\end{document}